\documentclass[a4paper]{article}
\usepackage{amssymb}
\usepackage{euscript}
\usepackage{graphicx}
\usepackage{color}
\usepackage{epsfig}
\usepackage{fullpage}

\newcommand{\image}[3]{
\begin{figure}[#1]
\begin{center}
\includegraphics{full_#2.eps}
\caption{\small#3}
\label{image:#2}
\end{center}
\end{figure}
}

\def\nVsHz{%
$\mbox{nV}/\sqrt{\mbox{Hz}}$}

\def\fAsHz{%
$\mbox{fA}/\sqrt{\mbox{Hz}}$}

\begin{document}
\title{Cryogenic differential  amplifier for NMR applications}

\author {
  V.~V.~Zavjalov\/\thanks{e-mail: vladislav.zavyalov@aalto.fi},
  A.~M.~Savin
  P.~J.~Hakonen\\
  \it\small
  Low Temperature Laboratory,
  Department of Applied Physics, Aalto University,\\
  \it\small
  PO Box 15100, FI-00076 AALTO, Finland
}


\date{\today}
\maketitle

\begin{abstract}
We have designed and characterized  a cryogenic amplifier for
use in $^3$He NMR spectrometry. The amplifier, with a power consumption
of  $\sim 2.5$~mW, works at temperatures down to 4~K. It has a
high-impedance input for measuring a signal from NMR resonant circuit, and
a 50~$\mathrm{\Omega}$ differential input which can be used for pick-up
compensation and gain calibration. At 4.2~K, the amplifier has a voltage
gain of 45, output resistance 146~$\mathrm{\Omega}$ and a 4.4~MHz
bandwidth starting from DC. At 1~MHz, the voltage and current noise
amount to 1.3~\nVsHz{}  and 12~\fAsHz{}, respectively, which yields an
optimal source impedance of $\sim 100$~k$\mathrm{\Omega}$.
\end{abstract}

\image{h!}{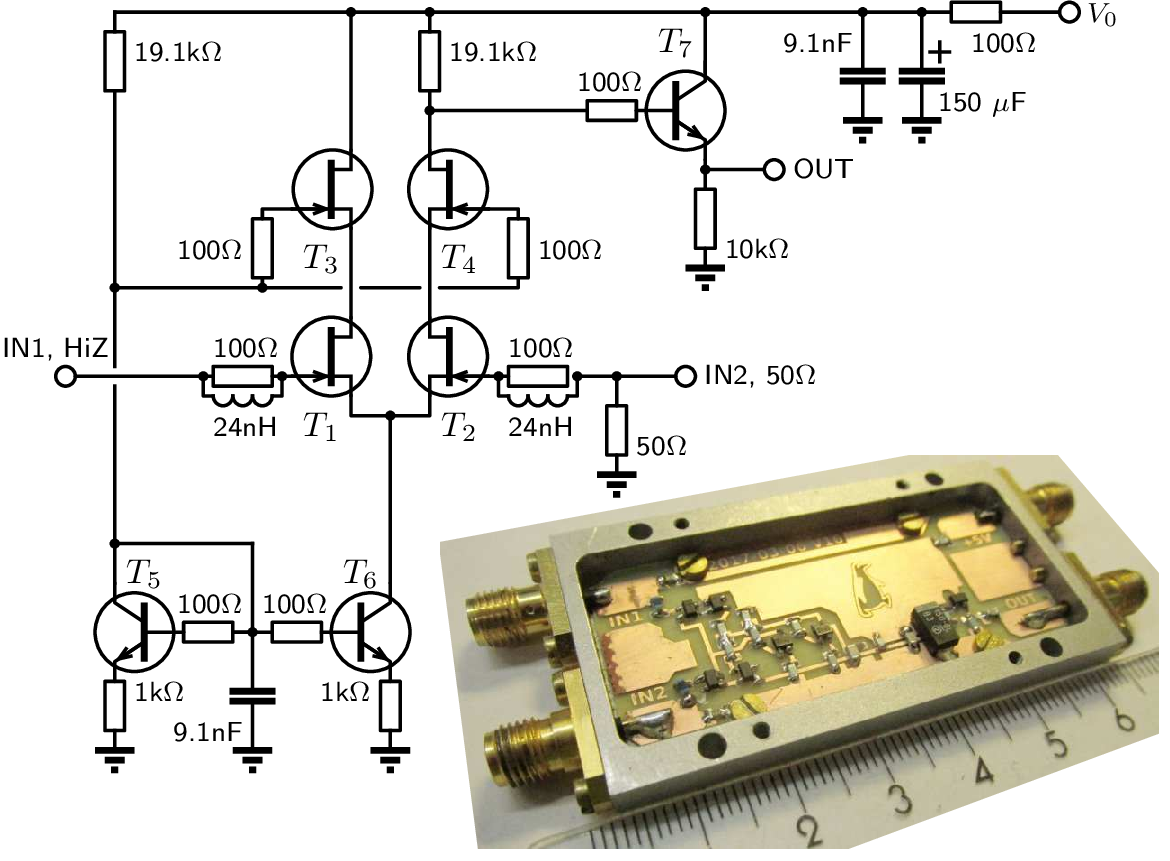}{Schematics and a photo of component layout.}

\section{Introduction}

Many scientific and research applications benefit from the use of
cryogenically cooled amplifiers. Reduction of the amplifier operation
temperature leads to significant improvement of the signal-to-noise
ratio, which finally improves sensitivity of measurements.

Nowadays, almost all commercially available transistors and operational
amplifiers use silicon technology, and they do not work below 10~K. On
the other hand most cryostats have a 4~K plate or liquid-helium bath
where cold amplifiers can be placed. With silicon-based amplifiers,
overheating with temperature stabilization can be
used~\cite{2009_hayashi} for normal operation. Many non-silicon
transistors used previously for cold amplifiers~\cite{1985_bloyet} are
not available anymore. Now there are a few types of low-noise non-silicon
transistors produced commercially which are designed for GHz frequencies.
They include GaAs HEMT-technology FET transistors made by Avago (ATF
series), CEL (CE series) or TriQuint (TGF series) and SiGe:C npn bipolar
transistors made by Infineon (BFP series). These transistors can be used
not only at high
frequencies~\cite{2005_pospieszalski,2004_roschier,2005_hu} but also for
low frequency
applications~\cite{2000_koivuniemi,2004_oukhanski,2004_robinson,2006_dicarlo,2007_vink,2008_mathur,2012_beev,2013_arakawa,2017_hirata}.
The performance of cooled low-frequency  amplifiers has also been
promoted by special, foundry-made transistors~\cite{2014_dong}.

In this paper, we present a cryogenic amplifier based on easily-available, commercial Avago ATF-33143 HEMT components
and Infineon BFP640 npn transistors. The goal has been to design 
a preamplifier for NMR experiments which operate around 1 MHz 
 and utilize tank circuits with high impedance ($\sim 100$~k$\mathrm{\Omega}$).

\section{Design}

The schematics and a photograph of the amplifier are shown in Fig.~1. The
differential cascode input stage consists of four HEMT transistors
$T_1-T_4$. The stage is powered by a current supply made of two npn
transistors $T_5-T_6$. The current is set by the bias voltage $V_0$ and a
19.1~k$\mathrm{\Omega}$ resistor (Note that base-emitter voltage of $T_5$ and $T_6$ is
close to 0.75~V). Cascode design together with non-symmetric
inputs are used to reduce the Miller effect~\cite{miller}. This effect takes
place when the output voltage couples to the input through transistor's
parasitic capacitance, and then, after amplification, decreases the total
gain. A 50~$\mathrm{\Omega}$ resistor on input~2 forms a divider with the
parasitic capacitance which reduces the effect. In equilibrium, the
current in both arms is half of the total current, and the output of
the amplifier is biased near $V_0/2$. A follower with npn
transistor $T_7$ is used to reduce the output resistance of the
amplifier. All transistors have 100~$\mathrm{\Omega}$ resistors connected to their
bases/gates. These resistors are needed to damp high-frequency resonances
caused by parasitic inductances of the circuit and to improve the stability of the amplifier. 
At the input terminals, the 100~$\mathrm{\Omega}$ resistors are shunted
by 24~nH inductors at the relevant signal frequencies. The bias voltage
line is connected through an RC-filter with a 100~$\mathrm{\Omega}$ resistor, a
150~$\mu$F tantalum capacitor and a 9.1~nF ceramic capacitor.

For the input stage we are using Avago ATF-33143 HEMTs. We have tested
transistors with smaller gate areas (ATF-34143, ATF-35143)  and obtained
similar results. For the current supply and the output follower we use
Infineon BFP640 npn SiGe transistors. Passive components for cryogenic
applications need to demonstrate both mechanical and electrical stability
at low temperatures. We are using ceramic C0G
capacitors~\cite{2010_teyssandier}, tantalum
capacitors~\cite{2006_teverovsky}, and metal-film (thin-film) resistors.

With the bias voltage $V_0=4$~V, the current source gives $\sim 0.2$~mA
and the total power consumption of the device amounts to~$\sim 2.5$~mW.
Low power consumption has certain disadvantages, namely high output
resistance and low slew rate. This makes the amplifier performance to be
dependent on the output circuit. If the amplifier is connected to an
output line with capacitance, two effects are observed at high
frequencies: the gain drops because the line capacitance forms an
RC-filter with the output resistance; at high amplitudes the signal is
distorted because the output current of the amplifier is limited and this
restricts the charging rate of the line capacitance. The second effect
can be estimated using the known parameters of the amplifier: the current
limit is determined by the bias voltage and the resistor in the output
emitter follower, $I_{\mbox{max}} \sim 0.4$~mA. The charging rate is
constrained by~$\dot V = 2\pi f\,V = I_{\mbox{max}} / C_{\mbox{line}}$.
For~$C_{\mbox{line}}=500$~pF and at~$f=1$~MHz, the distortion starts at
an output amplitude of~$V\sim 130$~mV. This estimation agrees with our
measurements. Using shorter output cables or adding an additional, more
powerful amplifier on the upper part of the cryostat would improve the
amplifier performance.

\section{Measurements}

To obtain output parameters of the amplifier (gain, bandwidth and the
output resistance), we measured the gain as a function of frequency for a
few different configurations: amplifier connected directly to a
measurement device (lock-in amplifier); 1~nF capacitor added between the
amplifier and the lock-in; a 60~$\mathrm{\Omega}$ resistor to the ground
added after the capacitor. All the data were fitted simultaneously with a
simple electrical model with four parameters: gain, bandwidth, the output
resistance and the cable capacitance. Measurements were done at room
temperature (300~K), temperature of liquid nitrogen (77~K) and liquid
helium (4.2~K). Typical cable capacitance in our setup was $\sim580$~pF
and temperature-independent. Other parameters are presented in
Table~\ref{table}. In all measurements presented here we used input~1
while input~2 was shunted to ground by 50~$\mathrm{\Omega}$ resistor.
With swapped input connections, we obtained very similar gain
characteristics as expected for a differential amplifier. The measured
gain of the amplifier (without the additional capacitor and grounding
resistor on the output) is plotted as a function of bias voltage~$V_0$
(at 1~MHz) and frequency (at~$V_0=4$~V) in Fig.~2a and~2b, respectively.
The measured cut-off frequency~$\sim 1$~MHz in Fig.~2b was determined
mostly by capacitance of the output cable which formed an RC filter with
the output resistance of the amplifier. The obtained intrinsic gain of
the amplifier extends over DC$\ldots$4~MHz and it is illustrated in
Fig.~2b by dashed curves. In Fig.2c we present a DC response of the
amplifier ($U_{\mbox{out}}/U_{\mbox{in}}$). In~Table~\ref{table} we
provide 1dB compression ranges obtained from this measurement. They show
input voltage ranges where the response deviates from linear function by
less then 1dB (in power units).

\image{h}{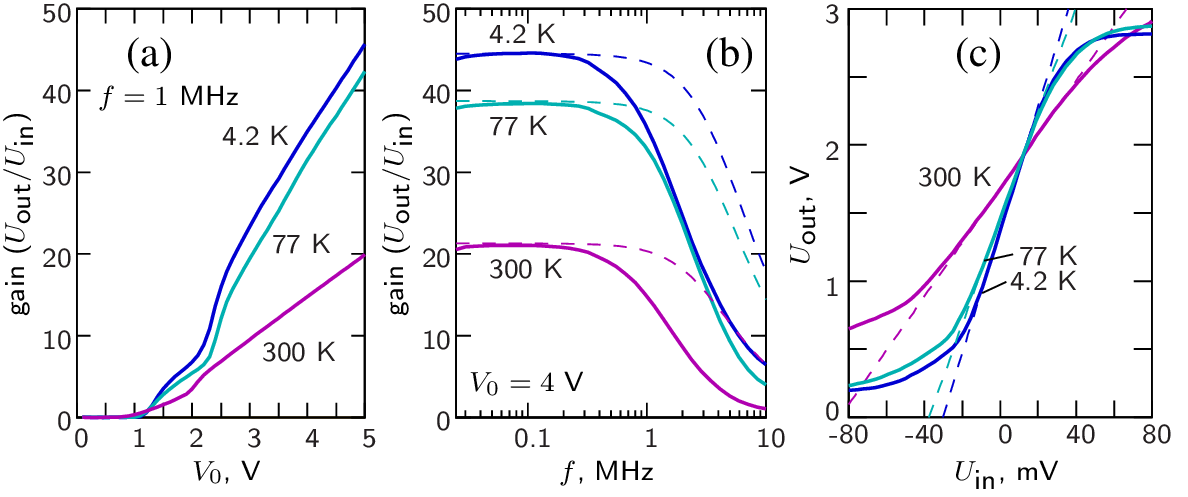}{ {\bf (a)} Measured AC gain at 1~MHz as a function of the
bias voltage~$V_0$. {\bf (b)} Measured AC gain at $V_0 = 4$~V as a
function of frequency. The 1~MHz cut-off frequency here is determined by
the output resistance of the amplifier ($\sim$150~$\mathrm{\Omega}$) and
the capacitance of output cables ($\sim$580~pF). The obtained intrinsic
bandwidth~$\sim4$~MHz is shown by dashed lines. {\bf (c)} DC response of
the amplifier ($U_{\mbox{out}}/U_{\mbox{in}}$). Measurements have been done
at $T=300$~K, 77~K, and 4.2~K.}

\image{h}{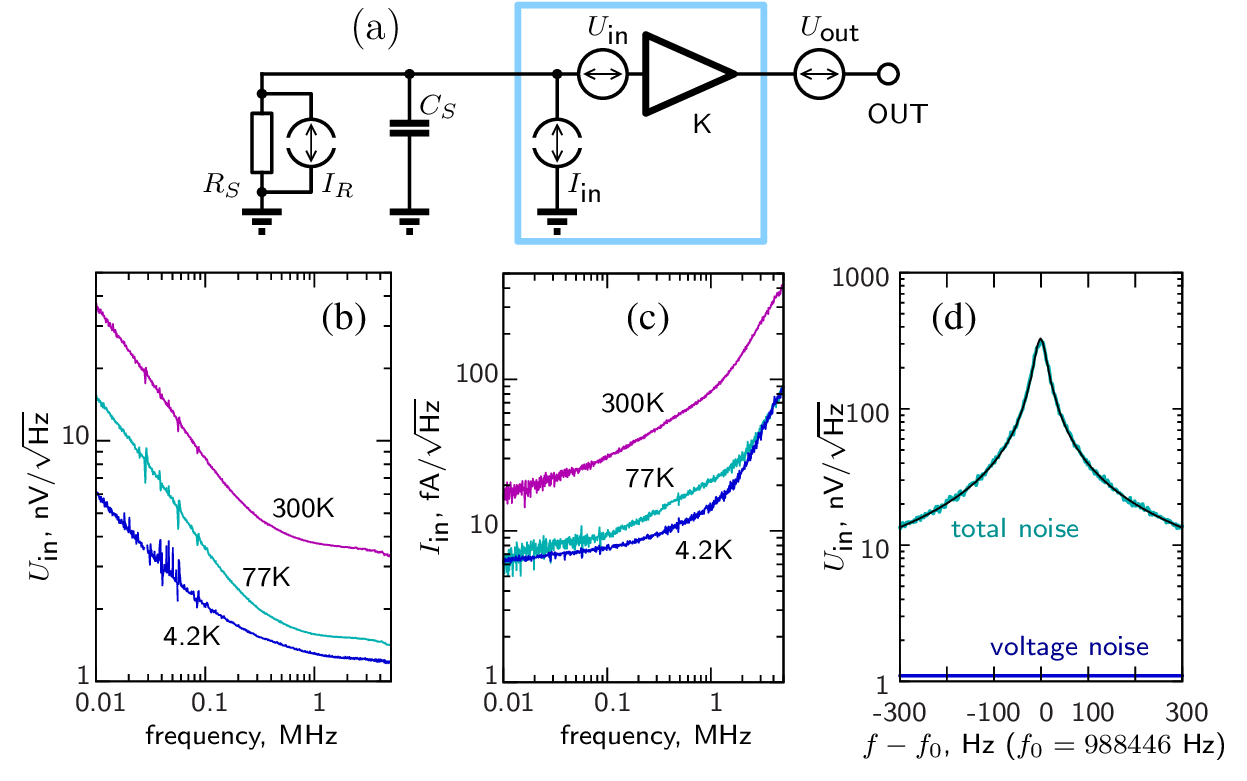}{ Model for noise measurements {\bf (a)}, measured
voltage {\bf (b)} and current {\bf (c)} noise of the amplifier.
Measurements have been done at 300~K, 77~K and 4.2~K using bias
$V_0=4$~V. {\bf (c)} Noise measured using a high-$Q$ tank circuit at the
input of the amplifier. The curve is a sum of small input voltage noise
1.3~nV/$\sqrt{\mbox{Hz}}$ and input current noise
11.8~fA/$\sqrt{\mbox{Hz}}$ imposed on the impedance of the tank circuit.}

A model used for our noise measurement analysis is shown in Fig.~3a. The
amplifier consists of a noiseless amplifier with voltage gain~$K$,
input voltage noise source $U_{\mbox{in}}$, and input current noise
source~$I_{\mbox{in}}$. The input of the amplifier is connected to a
resistor~$R_S$ which also works as a source of thermal noise~$I_R^2 =
4k_BT/R$, where $T$ is the temperature of the resistor. Capacitance~$C_S$
of the input circuit is also taken into account. Output of the amplifier
is connected to a digital oscilloscope which is used to record noise
spectra. The noise source~$U_{\mbox{out}}$ represents the noise of the
oscilloscope.

If the noise sources are not correlated, we can make a square sum of average
noise voltages produced by each source. Then the square of the measured
noise spectral density becomes
\begin{equation}\nonumber
U^2 = U_{\mbox{out}}^2 + K^2 \left[ U_{\mbox{in}}^2
   + \left(I_{\mbox{in}}^2 + I_R^2\right) |Z_S|^2 \right],
\end{equation}
$$
|Z_S|^2 = \frac{R_S^2}{1+(\omega R_S C_S)^2}.
$$
Here $Z_S$ is the total impedance of the source ($R_S$ and $C_S$ in parallel)
and~$\mathrm{\omega}$ denotes the angular frequency.

It is easy to find~$U_{\mbox{out}}$ and~$U_{\mbox{in}}$ by measuring two
noise spectra: one with the amplifier powered off ($K=0$) and another
with shorted input ($R_S=0$). For finding~$U_{\mbox{in}}$, one have to
subtract one spectrum from another and divide square root of the result
by known gain.

To determine the input current noise~$I_{\mbox{in}}$, we recorded noise
spectra with non-zero source resistance~$R_S$. The noise with~$R_S=0$ was
subtracted (removing the effects of $U_{\mbox{out}}$ and~$U_{\mbox{in}}$)
and the result was converted to amplifier input. This gives us a
source-dependent component of the input noise, $(I_{\mbox{in}}^2 + I_R^2)
|Z_S|^2$. For small~$R_S$, the main contribution is thermal noise. A
cut-off frequency $1/R_S C_S$ is clearly seen on the corresponding noise
spectra, which allowed us to obtain an estimate for the capacitance at
the input $C_S=6.7$~pF. The input capacitance of the transistor itself is
near 1~pF, the rest is determined mostly by SMA connector and the PCB
design. Knowing $C_S$, we evaluated the impedance~$Z_S(\omega)$ and
extracted the current noise. We did this using
$R_S=10$~M$\mathrm{\Omega}$ at which the thermal current noise $I_R$ is
smaller. The obtained values for~$U_{\mbox{in}}$ and~$I_{\mbox{in}}$ are
displayed in Fig.~3b and~3c.

The voltage noise determination described above was well defined and
straightforward, but the measurement of the current noise was more tricky:
a large thermal noise part had to be subtracted from the result.
To decrease the thermal noise $I_R$ one has to use huge source resistances,
but this limits the bandwidth because of the $RC$-filtering by the input
capacitance of the amplifier. In addition, the parasitic input resistance may
affect the result if it is comparable with the source resistance.

In order to improve the accuracy of the current-noise measurements, we
have made another experiment utilizing a resonant circuit as a high-impedance
source, similarly to the work in Ref.~\cite{2000_koivuniemi}. The circuit
was formed by a niobium coil with $L=100$~$\mu$H and a capacitor
with~$C=200$~pF to obtain a resonance frequency near 1~MHz. We used an
American Technical Ceramics porcelain capacitor (100B series) which has
very low dissipation. Any resistive metal near the coil leads to extra
dissipation as well as thermal noise. To avoid this, we used a
superconducting wire without normal-metal matrix. The circuit was placed
in a solder-covered copper box which worked as a superconducting
enclosure. We concluded that if a small piece of normal metal inside the
enclose does not decrease the $Q$-value below a few thousand, then it has
no substantial effect on the noise.  This confirms that the measured
noise is originated from the amplifier.

The output voltage noise was recorded as described above. The gain of the
amplifier was calibrated using an external noise source connected to
input~2. Near 1~MHz, the equivalent input noise contains a constant
voltage noise ($\approx1.3$~nV/$\sqrt{\mbox{Hz}}$) and a
Lorentzian-shaped noise peak, produced by current noise imposed on the
impedance of the tank circuit.

With all the precautions, we managed to reach $Q\approx1.2\cdot10^6$,
which corresponds to the maximum impedance $Z=0.9$~G$\mathrm{\Omega}$ and
an effective dissipative resistance 0.5~m$\mathrm{\Omega}$ in series with
the coil. This confirms that the input impedance of the amplifier is
higher than~$0.9$~G$\mathrm{\Omega}$. It is hard to make good noise
measurements with such a high $Q$-value circuit because of the long time
needed for good frequency resolution and frequency drifts during this
time. For avoiding these problems in our tank-circuit noise measurements, 
we added a small resistor to reduce the $Q$-value to $40\cdot10^3$
($Z=28$~M$\mathrm{\Omega}$). The noise curve measured with such a circuit
is depicted in Fig.~3d. According to the obtained data, the current noise
of the amplifier is 11.8~fA/$\sqrt{\mbox{Hz}}$, which is in a good
agreement with the above-described measurements with a
10~M$\mathrm{\Omega}$ resistive source.

\begin{table}
\begin{tabular}{|l|c|c|c|}\hline
Temperature $T$,~K           & 300  & 77   & 4.2 \\\hline
Gain $K_0$ ($U_{\mbox{out}}$/$U_{\mbox{in}}$)& 21.3 & 38.7 & 44.5 \\
$R_{\mbox{out}}$,~$\mathrm{\Omega}$   & 233  & 128  & 146  \\
Cut-off frequency $f_c$, MHz & 3.23 & 4.02 & 4.41\\
1dB compression range, mV & $-20.6 \ldots 35.6$
& $-7.93\ldots14.4$
& $-5.13\ldots13.2$\\
[3pt]
Voltage noise, nV/$\sqrt{\mbox{Hz}}:$&&&\\
at 10~kHz                    & 35.2 & 14.9 & 6.0 \\
at 100~kHz                   &  8.4 &  3.6 & 2.1 \\
at 1~MHz                     &  3.8 &  1.6 & 1.3 \\
[3pt]
Current noise, fA/$\sqrt{\mbox{Hz}}:$&&&\\
at 10~kHz                    & 17.5 &  6.2 &  6.4 \\
at 100~kHz                   & 30.9 &  9.3 &  7.7 \\
at 1~MHz                     & 83.3 & 21.4 & 14.5 \\
[3pt]
Optimal source impedance, $\mathrm{\Omega}$:&&&\\
at 10~kHz                    & 2.0M & 2.4M & 930k \\
at 100~kHz                   & 270k & 390k & 270k \\
at 1~MHz                     &  45k &  74k &  89k \\
\hline
\end{tabular}
\caption{Characteristics of the amplifier measured at three different
temperatures. Bias voltage~$V_0=4$~V has been  employed in all
measurements. Current noise at all frequencies was measured with resistive
source. More accurate measurement with a resonant circuit
gave~11.8~fA/$\sqrt{\mbox{Hz}}$ at 1~MHz.}
\label{table}
\end{table}

We made and tested a few similar amplifiers. Altogether, it was easy to make a
working device and to get the expected characteristics. The gain is determined by
the input HEMT transistors, the characteristics of which may vary. We observed about 15\% scatter in
the gain value between different devices. The balancing of the transistors is
also important: one has to check that the DC response is symmetric and
replace HEMT transistors if needed. We did not observe any noticeable
change of the amplifier characteristics after $10-20$ fast cooling and heating
cycles. Floating of input~1 can cause electrostatic damage to the input HEMT
transistor. To avoid this one can ground input~1 by a 1~M$\mathrm{\Omega}$ resistor,
but depending on the impedance of the signal source, the noise contribution of
this resistor can be significant. Noise coming from input~2 can increase
both the current and voltage noise of the amplifier. One has to incorporate a sufficient amount 
attenuation at low $T$ when connecting input~2 to room-temperature devices.

\image{h!}{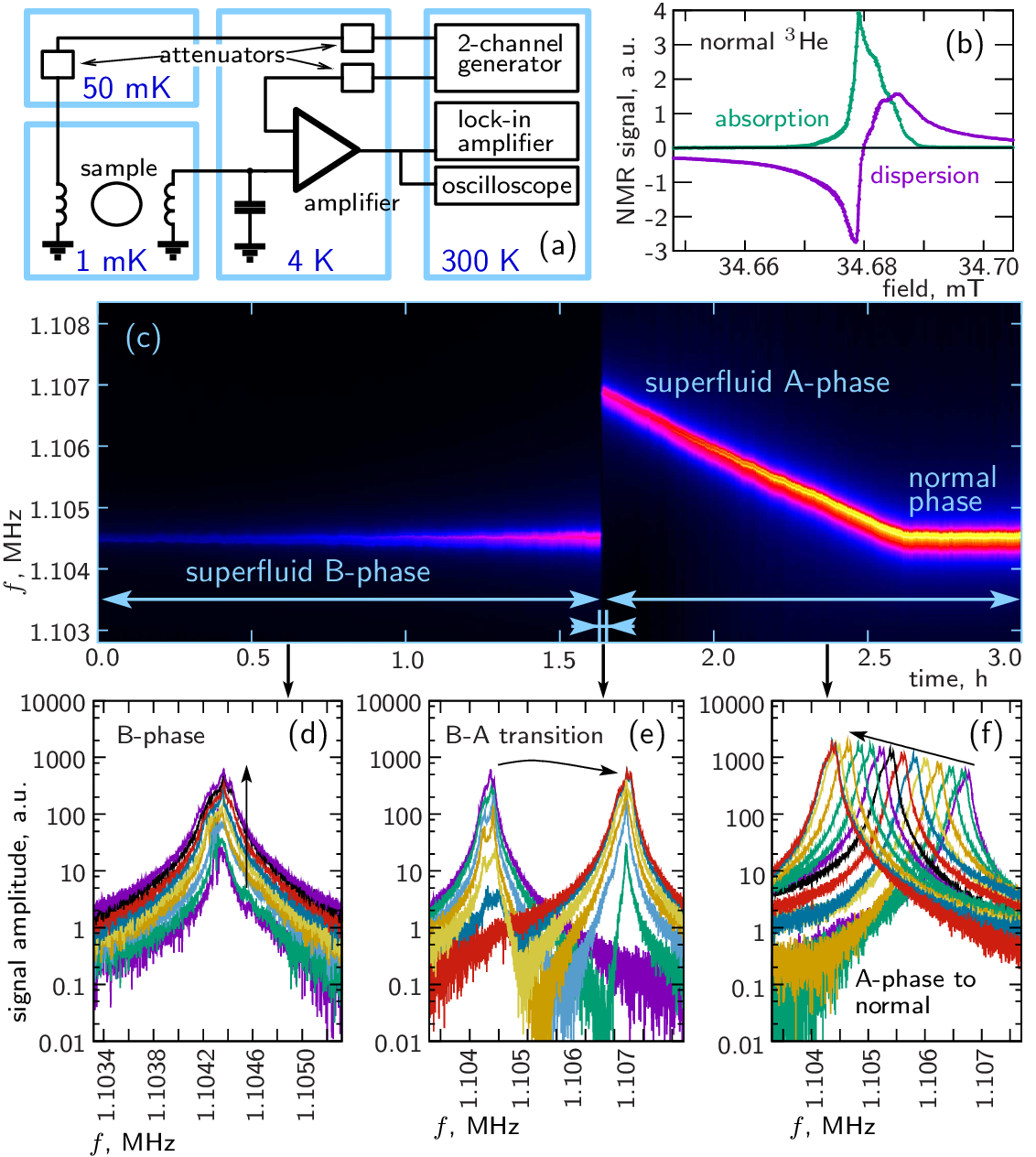}{Application example. {\bf (a)} Measurement setup (see
text). {\bf (b)} Typical continuous-wave NMR signal in normal $^3$He at
pressure~$29.5$~bar, temperature~$T=20$~mK, and frequency~$f=1.10$~MHz.
Base level of the signal is near zero because of the compensation imposed
to the second input of the amplifier. {\bf (c)} Pulsed NMR measurement
during warming up at the same pressure and NMR frequency. Two phase
transitions can be seen when $^3$He traverses through B and A superfluid
phases and warms up to the normal phase. Each vertical slice of the
picture is a power spectrum of a signal recorded by the oscilloscope
after a short excitation pulse from the generator. Examples of such
spectra are shown below: {\bf (d)} in the B-phase, {\bf (e)} during B
to A transition, and {\bf (e)} in A and normal phases. Black arrows
on the plots show evolution of the signal during warming up.}

\section{Application example}

We have used the amplifier for NMR experiments in superfluid~$^3$He.
Schematics of the NMR spectrometer is shown in Fig.~4a. The $^3$He sample
is located in a magnetic field of about $35$~mT. A pair of crossed coils
is used to apply a radio-frequency excitation from a generator and to
pick up a signal from precessing magnetization. The pick-up coil
($L\sim50\mu$H) together with a capacitor forms a tank circuit with a
resonant frequency $1$~MHz, $Q \sim 200$ and an on-resonance
resistance of $\sim 70$~k$\mathrm{\Omega}$. This source resistance is
close to the optimal impedance of the amplifier, which means that effects
of voltage and current noise are comparable. Using the values from
Table~1, one obtains for the total input noise~2.3~nV/$\sqrt{\mbox{Hz}}$.
Even outside the nuclear magnetic resonance condition, there is a coupled
signal on the amplifier due to the mutual inductance and capacitance
between the excitation and pick-up coils. We are using the second input
of the amplifier to compensate for this signal. This compensation is not
affected by possible drifts of the amplifier gain and can be used also
for calibration of the NMR signal amplitude. Output of the amplifier is
connected to a lock-in amplifier and to an oscilloscope.

In Fig.~4b a typical continuous-wave NMR signal in normal $^3$He is shown
to provide a qualitative understanding of signal-to-noise ratio in our
measurements. Figures~4c-4f demonstrate another type of measurement,
pulsed NMR. A short excitation pulse is used to excite precession of
magnetization, and free-induction decay is recorded by oscilloscope. In
Fig.~4c spectra of such signals are presented as a color density plot
with time and frequency coordinates. The measurement have been done
during warming up from superfluid B-phase of~$^3$He, through superfluid
A-phase to the normal phase. Both transitions and all the usual features
such as change of magnetic susceptibility and temperature-dependent
frequency shift in the A-phase are seen. In Fig.~4d-4f individual spectra
(slices of the color plot) are shown.

\section{Conclusions}

A differential cryogenic amplifier has been developed and used for
high-resolution NMR measurements. The amplifier consumes only 2.5~mW of
power, which facilitates installation of many such amplifiers on a 4~K
stage of a regular dilution refrigerator. The input current noise of the
amplifier is 12~fA/$\sqrt{\mbox{Hz}}$ at 1~MHz, which renders our
amplifier excellent for high-$Q$, tuned NMR probes up to impedances of a
few hundred k$\mathrm{\Omega}$.  For the intended NMR applications, it is also
important to be able to compensate for the cross-talk input signal on the
first amplifier stage, and to measure the amplifier gain in-situ using
the differential input. Our design can be operated even at a smaller
power consumption level, but with a slight loss in the amplifier gain. In
future, after optimization for lower bias voltage operation, we plan to
position the amplifier on the still plate of a
dilution refrigerator, which would further help in minimizing possible
external interferences and parasitic impedance problems in cryogenic
input circuitry.

This work was performed as part of the ERC QuDeT project (670743). It was
also supported by Academy of Finland grants 310086 (LTnoise) and 312295
(CoE QTF). This research project made use of the Aalto University
OtaNano/LTL infrastructure which is part of European Microkelvin
Platform.

A commercial version of the amplifier is available from Aivon OY~\cite{aivon}.

\end{document}